# IGR J17544-2619 in depth with *Suzaku*: direct evidence for clumpy winds in a supergiant fast X-ray transient

Rachel A. Rampy[1], David M. Smith[1], and Ignacio Negueruela[2]
(1) Physics Department and Santa Cruz Institute for Particle Physics, University of California, Santa Cruz, 1156 High St., Santa Cruz, CA, USA. rrampy@ucsc.edu
(2) Departamento de Física, Ingeniería de Sistemas y Teoría de la Señal, Universidad de Alicante, Apartado 99 E03080, Alicante, Spain

Abstract

We present direct evidence for dense clumps of matter in the companion wind in a Supergiant Fast X-ray Transient (SFXT) binary. This is seen as a brief period of enhanced absorption during one of the bright, fast flares that distinguish these systems. The object under study was IGR J17544-2619, and a total of 236 ks of data were accumulated with the Japanese satellite *Suzaku*. The activity in this period spans a dynamic range of almost $10^4$ in luminosity and gives a detailed look at SFXT behavior.

## 1. INTRODUCTION

The last decade has seen the discovery of two new classes of high-mass X-ray binaries (HMXBs), some with very high absorption (Walter et al. 2006), and some, featuring brief transient outbursts, called Supergiant Fast X-ray Transients (SFXTs) (Negueruela et al. 2005). The first fast transient identified with a supergiant companion was XTE J1739-302 (Smith et al. 2003), and data from INTEGRAL and other observatories quickly made it clear that there are many such systems (in't Zand 2005; Sguera et al. 2005; Negueruela et al. 2005). SFXTs have spectra similar to persistent supergiant HMXBs, but only occasionally and briefly reach comparable luminosities. Proposed mechanisms responsible for this behavior are clumping of the wind from the supergiant star (in't Zand 2005; Walter & Zurita Heras, 2007; Negueruela et al., 2008), the presence of a dense equatorial wind component (Sidoli et al. 2007), and gated accretion due to the propeller effect from the magnetosphere of the compact object (Grebenev and Sunyaev 2007; Bozzo et al. 2008).

IGR J17544-2619 is an archetypical SFXT, with a history of rapid, seemingly sporadic outbursts of strongly variable flux (e.g. Gonzalez-Riestra et al. 2004; Krimm et al. 2007; Kuulkers et al. 2007). It was discovered in 2003 (Sunyaev et al. 2003), is associated with an O9Ib supergiant located at ~3.6 kpc (Pellizza, Chaty and Negueruela 2006, Rahoui et al. 2008) and probably harbors a neutron star (in't Zand 2005). Suggestions of a long periodicity in the outbursts were made by Walter et al. (2006; a period of 165 ± 3 dy) and Sidoli et al. (2009; a period of ~150 dy), but a recent periodogram analysis of the overall outburst history by Clark et al. (2009) found a strong periodic signal at a period of 4.926 ± 0.001 dy that is presumably the true orbital period. Other SFXTs have recently been shown to havea remarkable range of periodicities: IGR J11215-5952 (a strict periodicity at around 165 days; Sidoli et al. 2007), SAX J1818.6-1703 (an active phase of about 6 days occurring every ~30 days; Bird et al. 2009, Zurita Heras & Chaty 2009), and IGR J16479-4514 (3.3 days; Jain et al. 2009). IGR J17544-2619 and IGR J16479-4514 thus have periods comparable to those of persistent supergiant HMXBs.

*Suzaku* is Japan's fifth X-ray astronomy satellite. It was developed under Japan-US collaboration and launched on 2005 July 10 (Mitsuda et al. 2007). It allows high-sensitivity wide-band X-ray spectroscopy, with low and stable detector backgrounds. We present observations of IGR J17544-2619 with three of the X-ray imaging spectrometers (the XIS), which are sensitive to energies between 0.5 – 10 keV, and the silicon (PIN) layer of the Hard X-ray Detector (HXD) in the range 12 – 70 keV.

## 2. OBSERVATION AND DATA REDUCTION

The observation of IGR J17544-2619 took place from 2008 March 19 through March 22. The XIS were operated in 1/4 window mode, resulting in lightcurves with 2 s time resolution. For a small portion of the observation (~30 min), high count rates caused saturation to occur in the XIS instruments and contributed to data transfer problems in the HXD. To correct for this in the affected portion of XIS data, a circle of 15 pixel radius was excised from the center of the PSF. Recovery of HXD PIN data from this period was accomplished with intervention by the *Suzaku* user-support team. All other data reduction has been performed with HEAsoft 6.5, as directed in the *Suzaku* Data Reduction Guide[1].

For the XIS, background was taken from source-free regions near the edges of each CCD. Lightcurves and spectra were produced with `xselect`, and response files were generated with the tools `xisrmfgen` and `xissimarfgen`, which take into account extraction region, vignetting, etc. Spectra from the front-illuminated CCDs (XIS0 and XIS3) were combined and then fit simultaneously in XSPEC (version 12.4.0) with the back-illuminated data (XIS1), and (in one case) HXD PIN data as well. In all instances, the relative normalizations among the different types of instruments were allowed to vary freely. For the PIN spectra, "tuned" background files were used to account for the non-X-ray background. These are simulated event files generated by the HXD team after each observation. They are dead-time corrected and have estimated systematic uncertainties of 1.3%.

## 3. RESULTS

### 3.1 Lightcurves

Figure 1 shows the XIS count rate during the entire observation, summing the background-subtracted lightcurves from the three XIS detectors. The times are relative to a starting time at MJD = 54544.52698 or 2008 March at 12:38:51 UTC. In the ephemeris of Clark et al

1 See http://heasarc.gsfc.nasa.gov/docs/suzaku/analysis/abc/

(2009), in which the highest fluxes come at phase 0.0 relative to MJD 52702.9, our observation starts at a phase of 0.86 and ends at 0.41, covering a bit more than half of a complete orbit.

There is an intense period of flaring for ~1 day, with the flux during the brightest peak ~9000 times greater than in the first 10 hr of the observation. This occurs at hour 36 in Figure 1, which has a phase of 0.16 in the Clark et al. ephemeris. An isolated medium-sized flare is seen ~15 hr after the primary activity subsides, at 58 hr in Figure 1 (orbital phase 0.35). The most intense period of activity consists of multiple flares, with typical durations of ~3 min, and ends with the large flare at 36 hr (peak flux $3.4 \times 10^{-9}$ erg $cm^{-2}$ $s^{-1}$), which decays over a period of ~30 min. Figure 2 shows one *Suzaku* orbit from the most active period of flaring. Unfortunately, the rise of the biggest flare occurred while *Suzaku's* view was obstructed by the Earth.

The fastest time variability is seen in the rise of the isolated flare at 58 hr, which doubles in a remarkable 4 s (Figure 3). Bozzo et al. (2009) saw secondary flaring in IGR J16479-4514 and suggested the possibility of a stable structure in the supergiant wind. The recently published short orbital periods of both sources support this notion, since the features seen in Figure 1 cover substantial stretches of the orbit. We note that the large outburst at 36 hr falls exactly halfway between the first small outburst at 13 hr and the last outburst at 58 hr, suggesting that our observations might span a periastron that occurs at the brightest outburst. Further in-depth observations may be required to see if the difference between the peak activity phase of Clark et al. (2009) at 0.0 and our brightest outburst at phase 0.16 is due to random variations from orbit to orbit or to uncertainties in the ephemeris.

From the bottom panel of Figure 1, where the count rate is logarithmic, it is apparent that the flares are not completely isolated, but occur on pedestals of low-level emission. Such a pedestal is also seen in IGR J11215–5952, which has an active phase lasting on average ~8 days out of its 165-dy orbit (Sidoli et al. 2007). Passage through a dense equatorial outflow at the periastron of a highly eccentric orbit has been proposed to explain this phenomenon (Romano et al. 2009).

3.2 Spectroscopy

We first divided the XIS data into six time intervals, with each resulting spectrum fit with an absorbed power law in the energy range 1 – 10 keV. These intervals correspond to the initial low-level period (A), medium flaring (B), major flare, post flare period (C), isolated medium flare (D), and final low-level period (E). Figure 4 shows these divisions, and the resulting $N_H$ and $\Gamma$ (photon index), with the remaining fit parameters shown in Table 1. Also in Table 1, in the first row, is a subset of period A: the first 32 ksec of the observation, which isolates the period of lowest flux. As expected from the results of previous studies (Walter et al. 2006, Sidoli et al. 2008), we find progressive hardening during the first four epochs, from lowest to highest flux: $\Gamma$ decreases from ~ 2.06 to 0.96 with no significant change in absorbing column (all ~ $2 \times 10^{22}$ $N_H$ $cm^{-2}$). The isolated medium flare (D) has a photon index similar to the lowest level emission, but a significantly greater $N_H$ than in any of the other time periods ($2.7 \times 10^{22}$ $cm^{-2}$).

Coincidentally, one monitoring observation in a study being conducted with *Swift* occured during our observation, at the end of period A (2008 March 20, 12:57:18 – 13:07:58 UT; Sidoli et al. 2009). The spectral fit to this pointing was extremely close to our period A values in Table 1, although the *Swift* error bars are larger (L. Sidoli, private communication). In the preprint version of Clark et al. (2009), this coincidence between *Suzaku* and *Swift* is incorrectly given in their Table 1 as having taken place on March 31, resulting also in an incorrect orbital phase of 0.35 with their ephemeris.

Four more spectra were accumulated by summing all intervals that had count rates within designated ranges. These were chosen such that each interval contained approximately one quarter of the total number of photon counts. This translated to selecting portions of the observation with rates of 0 - 3, 4 – 40, 41 – 75, and 76 – 118 events per second, as measured by XIS 1. The results were spectra with exposure times of 99.3 ks, 4.2 ks, 740 s, and 520 s, respectively. Table 2 contains fit and model parameters. Here we also see trend of spectral hardening with increasing flux, but $N_H$ does not vary monotonically with flux.

For the orbit containing the major flare, spectroscopy on even finer timescales is possible. Figure 5 shows these divisions superimposed on lightcurves in the energy ranges 1 – 1.75 keV and 5.5 – 8 keV. The low-energy lightcurve shows dips not present in the high-energy time history, suggesting variable absorption. We divided this portion of the observation into six intervals, ranging from 120 to 820 s in duration, chosen to correspond to times where the lowest-energy lightcurve showed variability not present at higher energies. Table 3 lists the parameters of spectral fits with an absorbed powerlaw model. Here it is evident that there is dramatic variability in the absorption on very short timescales, while the index and unabsorbed flux remain unchanged. In particular, $N_H$ more than doubles within ~2 min, and the period of highest absorption lasts only ~5 min. Spectra from the first three intervals and the last interval are shown in Figure 6. The first three clearly show variation restricted only to the lowest energies, consistent with highly variable absorption on this short timescale. The comparatively stable values of $N_H$ for the longer integrations in Tables 1 and 2 may simply be the result of averaging over greater variability on short timescales. Sidoli et al. (2009) saw a similar rapid variability in absorption from XTE J1739-302, another canonical SFXT, using the *Swift* X-Ray Telescope (XRT).

To see whether spectral properties are different during the rise and fall of an outburst, we selected and summed the rising and falling parts of all the clear peaks we could identify (with the exception of the large outburst, for which only the decline was available). See Figure 2 for an example of the time structure. Table 4 contains the resulting spectral parameters. No significant spectral difference is seen between the rising and falling intervals. In addition, Figure 2 shows no consistent asymmetry in count rate such as a fast rise and exponential decay (although the single peak in Figure 3 might be interpreted this way). These symmetries confirm the expectation that wind material is accreted directly and not temporarily stored in a small accretion disk, which could produce time-profile and spectral differences between the rise and fall of an outburst, such as those seen on longer timescales in low-mass X-ray binaries with large disks (e.g. Miyamoto et al. 1995).

Finally, for the main flare and a period of moderate flaring (B), we fit the XIS spectra simultaneously with the HXD PIN spectrum. The models were a power law, a power law times an exponential decay, and thermal bremsstrahlung (`bremss` in XSPEC). For period (B), the simple power law by itself resulted in a poor fit, with reduced $\chi^2$ = 1.46 for 926 degrees of freedom (dof), and the bremsstrahlung model, while it fit much better ($\chi^2$ = 1.05, 926 dof), was systematically high from ~20 – 40 keV. The powerlaw with an exponential cutoff beginning at zero energy, which has one more free parameter, fit best ($\chi^2$ = 0.89, 925 dof) and is shown with residuals in Figure 7. Table 5 has the parameters for all three fits. The same spectral model was shown to be a good fit to the broadband spectrum of the SFXT XTE J1739-302 (Blay et al. 2008). The superiority of this model to the other two is even clearer in the fit to the decay of the large flare, which has better statistics. The simple power law and bremsstrahlung spectra had reduced $\chi^2$ of 12.5 and 7.0, respectively, while the cutoff power law had a much better reduced $\chi^2$ of 1.27, with $N_H = (1.69 \pm 0.02) \times 10^{22}$ cm$^{-2}$, power law index (0.295 ± 0.014) and folding energy (8.00 ± 0.12) keV. The spectrum and fit are shown in Figure 8. There is no evidence for cyclotron scattering features despite the excellent counting statistics.

The good statistics of the spectrum shown in Figure 8 also allow for a sensitive search for iron fluorescence and absorption seen in other highly variable, wind-accreting HMXBs. A fluorescence line was seen in IGR J16207-5129 by Tomsick et al. (2009) at an equivalent width of 42 +/- 12 eV. In the large flare from IGR J17544-2619 we find a much lower (and not significant) equivalent width: a best-fit value of 7.3 eV with a 95%-confidence upper limit of 14.2 eV. In this fit, the line was assumed to be narrow and the centroid was initially fixed at 6.39 keV (the value from Tomsick et al.) but then allowed to fit, converging at 6.41 +/- 0.05 keV. In AX J1845.0-0433, Zurita Heras et al. (2009) found an absorption edge consistent with highly ionized iron at 7.9 keV. We find no evidence of this feature in our spectrum. The best fit depth for an edge at this energy is zero, with a 95%-confidence limit of 0.026. When the energy of the edge is allowed to vary, it fails to converge on any meaningful value.

## 4. DISCUSSION

This in-depth look at prototypical SFXT behavior reconfirms the nature of this class, with faint emission punctuated by eruptions of hard X-rays and significant absorption that varies in time. IGR J17544-2619 displayed a dynamic range of nearly $10^4$, with the average flux during the first 10 hours at $4 \times 10^{13}$ erg cm$^{-2}$ s$^{-1}$ (2 – 10 keV) and a maximum of $4 \times 10^{-9}$ erg cm$^{-2}$ s$^{-1}$ during the peak of the brightest flare. For an object at 3.6 kpc, this translates to a luminosity range of $6 \times 10^{32}$ to $6 \times 10^{36}$ erg s$^{-1}$. In the simplest clumpy-wind picture, the X-ray luminosity scales as the accretion rate, which is in turn proportional to the local wind density.

The faintest interval we observed (the first 32 ksec,or the first row in Table 1) is about 30 times brighter than the quiescent level in this source observed by in't Zand (2005) with *Chandra*, which showed a soft spectrum consistent with either thermal emission from the neutron star or the emission of the atmosphere of the supergiant companion. Our lowest level, with its hard spectrum, suggests active accretion at a flux (and presumably density) of $10^{-4}$ of the peak during outburst. The bright outburst seen by in't Zand (2005) rosein about an hour from complete quiescence, while the bright outburst we see here is preceded by a day of complex, rising activity. It is clear that much more observing of SFXTs with sensitive X-ray instruments will be necessary before we can understand what behavior is typical and what is exceptional.

The lightcurves show that periods of activity are made up of multiple fast flares, which can also appear as isolated events. Spectroscopy shows the average column density remaining fairly constant when averaged over long intervals but varying dramatically on short timescales, while the photon index hardens with intensity, in agreement with previous findings (Walter et al. 2006, Sidoli et al. 2008).

We were fortunate to catch a bright outburst in this observation. If all major flares last on the order of an hour, then in our 65-hour observation there would be only a 10% chance of seeing one, based on the RXTE Galactic bulge scan data (Swank & Markwardt 2001), which show only 1 sample out of 650 having a flux above $5\times10^{-10}$ erg cm$^{-2}$ s$^{-1}$. The RXTE observations are brief "snapshots" taken about once every three days, but demonstrate that outbursts like our brightest do not occur on most orbits.

We have analyzed the short-term variability of absorption for the first time in this system. The second time interval during the major flare (number 2 in Figure 5) shows an increase of $\sim1.7\times10^{22}$ $N_H$ cm$^{-2}$ that lasts only 300 s before starting to decline again. Figure 6 shows that this episode corresponds to dramatic spectral evolution below 4 keV, which is best explained by a sudden rise in the amount of neutral absorbing material along the line of sight. It seems plausible that this is due to the neutron star (NS) passing behind a clump of dense stellar wind at some larger radius. Assuming a supergiant with R = 20 solar radii, M = 25 solar masses, and an orbital radius for the NS of 3R, the NS will have an orbital velocity of 280 km s$^{-1}$. Then, if the cloud is circular and has no tangential velocity, it must have a radius of 42000 km (transverse to the vector toward the supergiant) and a mass of $1.5\times10^{18}$ g. There is no constraint on the size in the radial direction. In the case of complete accretion of the clump and a 10% efficiency for the conversion of rest mass to X-rays, such a cloud could fuel a 5-minute outburst at a level of $4.5\times10^{35}$ erg s$^{-1}$, comparable to the largest flares we see exclusive of the very bright outburst. Estimates of the clump masses required for SFXT outbursts range from this level (Negueruela et al. 2008) to as high as $10^{23}$ g (Walter and Zurita Heras 2007), but some of this variation is due to the range of outburst intensities that need to be explained. We expect clumps to have a wide distribution of sizes, both *a priori* and because of the wide range of flare sizes observed, so this first clump observed in absorption may not be representative of a typical outburst. Indeed, if clumps have a falling distribution in size, the first to be observed is most likely to be from the small end of the range that could have been observed.

Instead of a foreground clump, accreting material being focused toward the NS could be responsible for the absorption. A calculation of the ionization state due to the local X-ray environment and spectral fits with absorption from ionized and neutral material may be able to rule out such a scenario. This possibility will be investigated in future work. Regardless of the location of the density enhancement, however, this is direct evidence for exactly the sort of clump predicted by the clumpy-wind models.

The presence of optically thick clumps has also been reported in the persistent HMXBs Cygnus X-1 and Vela X-1 (Feng & Cui 2002; Kreykenbohm et al. 2008). Occultation by these clumps can last from minutes to hours, and can correspond to an increase in photoelectric absorption greater than one order of magnitude. We have shown that, as in Cyg X-1, absorption when measured on a short timescale in IGR J17544-2619 is much more variable than what is observed when integrations are taken over hours or more.

Pointed observations with INTEGRAL and AGILE have now shown Vela X-1, thought of as a persistent counterpart to the SFXTs, to undergo fast flaring and brief off states (Kreykenbohm et al. 2008; Soffitta et al. 2008), revealing a continuum of behavior between SFXTs and the "persistent" systems. Other supergiant binaries show more frequent outbursts than IGR J17544-2619 and XTE J1739-302 but much more frequent quiet periods than Vela X-1 (e.g. Walter & Zurita Heras 2007, Rahoui & Chaty 2008). Until recently, the available evidence pointed to a framework where the main difference between SFXTs and other supergiant HMXBs is diverse orbital geometries (e.g. Negueruela et al. 2008; Romano et al. 2009). The latest results showing orbital periods < 5 days, however, tend to support the idea of intrinsic differences in the companion winds or in the neutron star magnetosphere (Bozzo et al. 2009). Continued monitoring of these objects by instruments with the level of sensitivity available to *Suzaku* will be necessary to characterize the range of wind clumping parameters and the dependence on orbital phase and to unify the description of wind-accreting HMXBs.

Acknowledgments

This research is partially supported by NASA grant NNX08AL35G and by the Spanish Ministerio de Ciencia e Innovación (MICINN) under grants AYA2008-06166-C03-03 and CSD2006-70. We thank the scheduling and user-support teams of the *Suzaku* mission for their assistance. DMS thanks Laurel Ruhlen for useful conversations. We are grateful to the anonymous referee for pointing out a number of relevant results in the recent literature.

**Table 1:** Absorbed power-law fits to XIS data for seven time intervals (see text)[a].

| *Time* | *Duration (ks)* | $N_H$ $(10^{22}\ cm^{-2})$ | $\Gamma$ | *Norm. at 1 keV* $ph\ keV^{-1} cm^{-2} s^{-1}$ | *Reduced* $\chi^2$ *(dof)* | *Flux* $(erg\ cm^{-2} s^{-1})$ |
|---|---|---|---|---|---|---|
| Start | 32.01 | 1.99 $\pm$ 0.91 | 2.06 $\pm$ 0.60 | 2.07/2.19x$10^{-4}$ | 0.633 (26) | 4.07/4.30 x$10^{-13}$ |
| A | 84.16 | 2.15 $\pm$ 0.15 | 1.97 $\pm$ 0.09 | 9.40/9.92 x$10^{-4}$ | 0.720 (184) | 2.11/2.23 x$10^{-12}$ |
| B | 45.18 | 2.14 $\pm$ 0.03 | 1.37 $\pm$ 0.02 | 8.14/8.20 x$10^{-3}$ | 0.944 (868) | 4.86/4.89 x$10^{-11}$ |
| Flare | 1.85 | 2.13 $\pm$ 0.02 | 0.96 $\pm$ 0.08 | 0.180/0.192 | 1.20 (863) | 2.21/2.36 x$10^{-9}$ |
| C | 63.04 | 2.14 $\pm$ 0.10 | 2.18 $\pm$ 0.07 | 2.11/2.16 x$10^{-3}$ | 0.681 (235) | 3.38/3.47 x$10^{-12}$ |
| D | 4.28 | 2.71 $\pm$ 0.17 | 2.17 $\pm$ 0.09 | 1.29/1.36 x$10^{-2}$ | 0.685 (185) | 2.02/2.12 x$10^{-11}$ |
| E | 25.22 | 1.95 $\pm$ 0.44 | 2.34 $\pm$ 0.31 | 1.01/1.06 x$10^{-3}$ | 0.488 (65) | 1.29/1.35 x$10^{-12}$ |

[a]*Normalizations are for the back illuminated (XIS1) and front illuminated (XIS0+XIS3) respectively, with the exponent applying to both. Fluxes are for the energy range 2 – 10 keV, uncorrected for absorption, and are given in the same format as the normalization.*

**Table 2:** XIS spectra sorted by count rate[a].

| *XIS rate* *(ct s$^{-1}$)* | $N_H$ *($10^{22}$ cm$^{-2}$)* | $\Gamma$ | *Norm. at 1 keV* *ph keV$^{-1}$cm$^{-2}$ s$^{-1}$* | *Reduced $\chi^2$* *(dof)* | *Flux* *(erg cm$^{-2}$ s$^{-1}$)* |
|---|---|---|---|---|---|
| 0-3 | 2.31 ± 0.03 | 1.74 ± 0.02 | 2.92/3.12 x10$^{-3}$ | 1.51 (315) | 9.35/9.98 x10$^{-12}$ |
| 3-40 | 2.02 ± 0.03 | 1.10 ± 0.02 | 2.81/3.19 x10$^{-2}$ | 1.08 (282) | 2.72/3.09 x10$^{-10}$ |
| 40-75 | 2.11 ± 0.03 | 0.95 ± 0.01 | 0.176/0.187 | 1.10 (376) | 2.18/2.32 x10$^{-9}$ |
| 75-118 | 2.25 ± 0.03 | 0.94 ± 0.01 | 0.275/0.296 | 1.09 (410) | 3.47/3.73 x10$^{-9}$ |

*[a]The count rate is per XIS detector. The spectral model and format of the results are the same as in Table 1.*

**Table 3:** XIS spectra during the decay of the major flare (see Figure 5).

| *Time* | *Duration* *(sec)* | $N_H$ *($10^{22}$ cm$^{-2}$)* | $\Gamma$ | *Norm. at 1 keV* *ph keV$^{-1}$cm$^{-2}$ s$^{-1}$* | *Reduced $\chi^2$* *(dof)* | *Flux* *(x10$^{-9}$ erg cm$^{-2}$ s$^{-1}$)* |
|---|---|---|---|---|---|---|
| **1** | 120 | 1.47 ± 0.05 | 0.95 ± 0.03 | 0.266/0.289 | 0.544 (236) | 3.41/3.71 |
| **2** | 229 | 3.33 ± 0.08 | 0.89 ± 0.02 | 0.256/0.275 | 0.776 (401) | 3.34/3.59 |
| **3** | 339 | 2.56 ± 0.04 | 1.05 ± 0.02 | 0.318/0.345 | 0.998 (382) | 3.23/3.51 |
| **4** | 130 | 2.63 ± .12 | 1.01 ± 0.05 | 0.177/0.190 | 0.503 (193) | 1.94/2.08 |
| **5** | 150 | 1.61 ± 0.06 | 1.02 ± 0.03 | 0.183/0.194 | 0.737 (262) | 2.07/2.20 |
| **6** | 820 | 1.91 ± 0.04 | 1.04 ± 0.02 | 0.119/0.123 | 0.832 (381) | 1.28/1.33 |

**Table 4:** XIS spectra summed over intervals when the flux was rising and falling.

| | $N_H$ *($10^{22}$ cm$^{-2}$)* | $\Gamma$ | *Norm. at 1 keV* *ph keV$^{-1}$cm$^{-2}$ s$^{-1}$* | *Reduced $\chi^2$* *(dof)* | *Flux* *(x10$^{-10}$ erg cm$^{-2}$ s$^{-1}$)* |
|---|---|---|---|---|---|
| Rise | 2.02 ± 0.11 | 1.08 ± 0.05 | 1.62/1.66 x10$^{-2}$ | 0.595 (142) | 1.63/1.67 |
| Fall | 1.91 ± 0.08 | 1.17 ± 0.04 | 1.60/1.63 x10$^{-2}$ | 0.618 (214) | 1.39/1.41 |

**Table 5:** Fits to combined XIS/PIN data for period B (see Figure 4).

| *XSPEC model* | $N_H$ *($10^{22}$ cm$^{-2}$)* | $\Gamma$ | *kT* *(keV)* | *folding energy* *(keV)* | *Reduced $\chi^2$* *(dof)* |
|---|---|---|---|---|---|
| wabs*pow | 2.31 ± 0.03 | 1.46 ± 0.02 | ... | ... | 1.46 (926) |
| wabs*bremss | 2.20 ± 0.02 | ... | 26.4 ± 1.3 | ... | 1.05 (926) |
| wabs*pow*highecut | 1.84 ± 0.04 | 0.88 ± 0.03 | ... | 10.5 ± 0.6 | 0.89 (925) |

**Figures:**

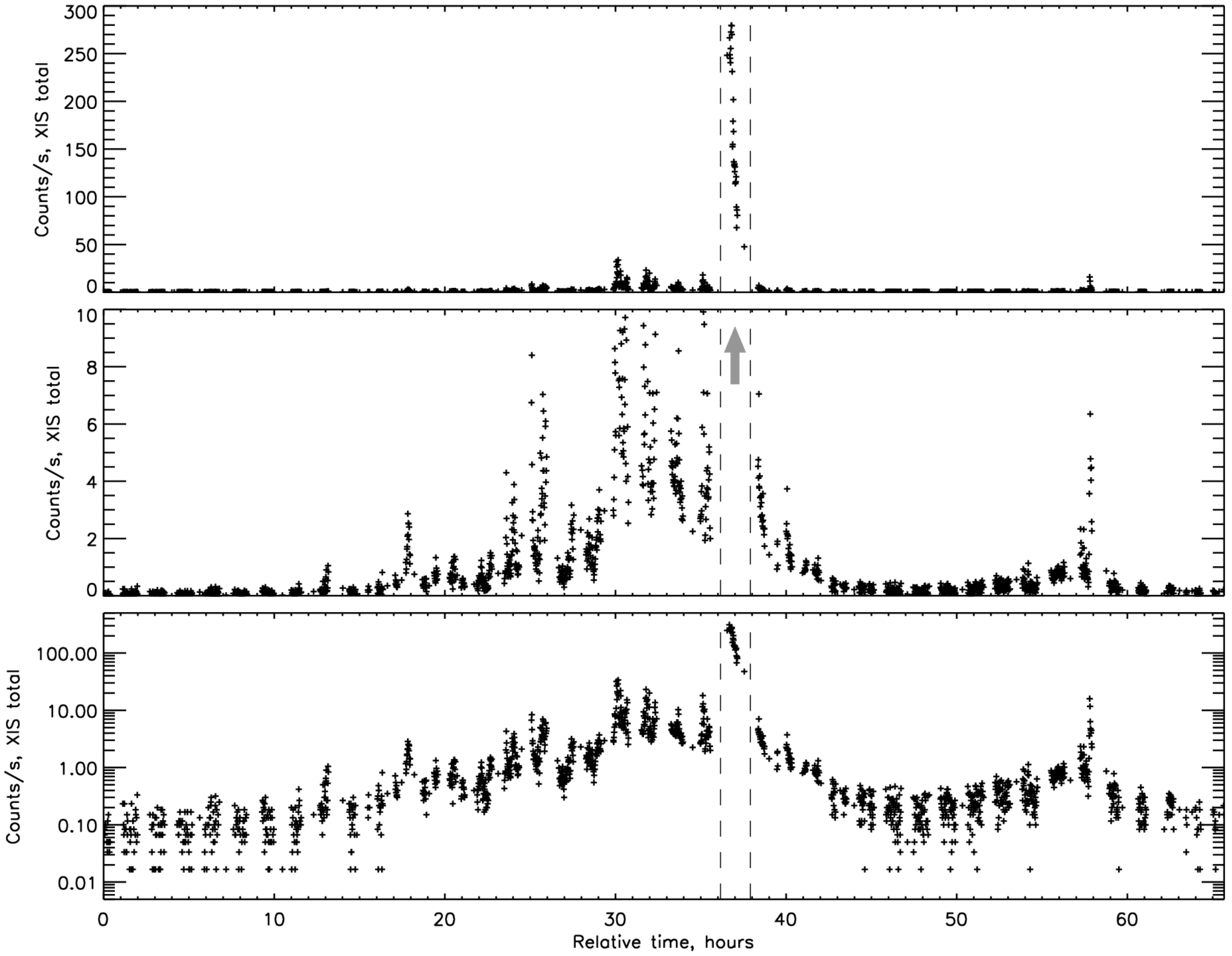

**Figure 1:** Lightcurves of the entire observation relative to MJD 54544.52698. Note the logarithmic scale in the bottom plot. The vertical dashed lines define the time period for the analysis shown in Figure 5, and the arrow in the center panel shows where the major outburst goes off scale.

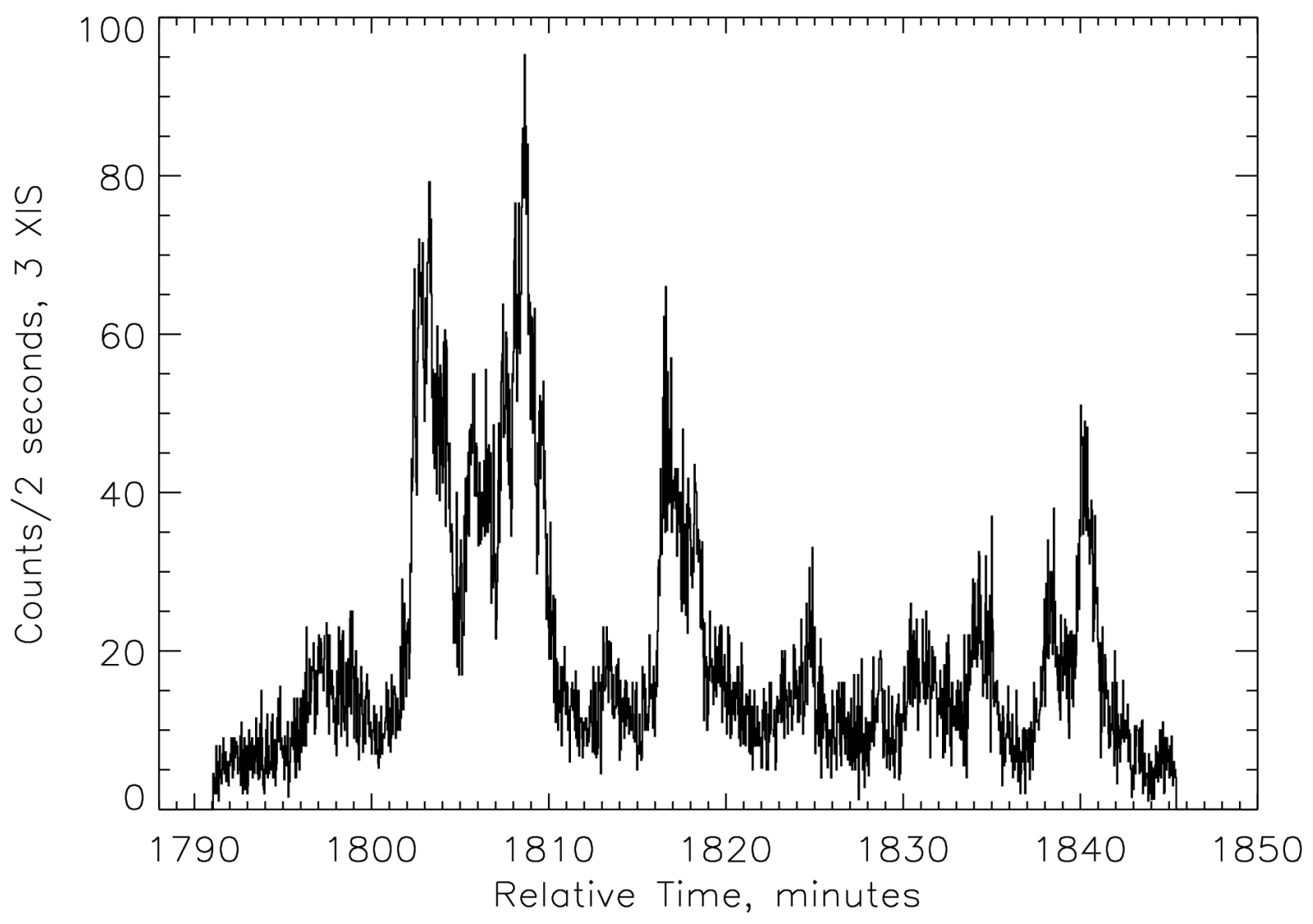

**Figure 2:** Lightcurve from one *Suzaku* orbit showing fast flares with average durations of ~3 minutes.

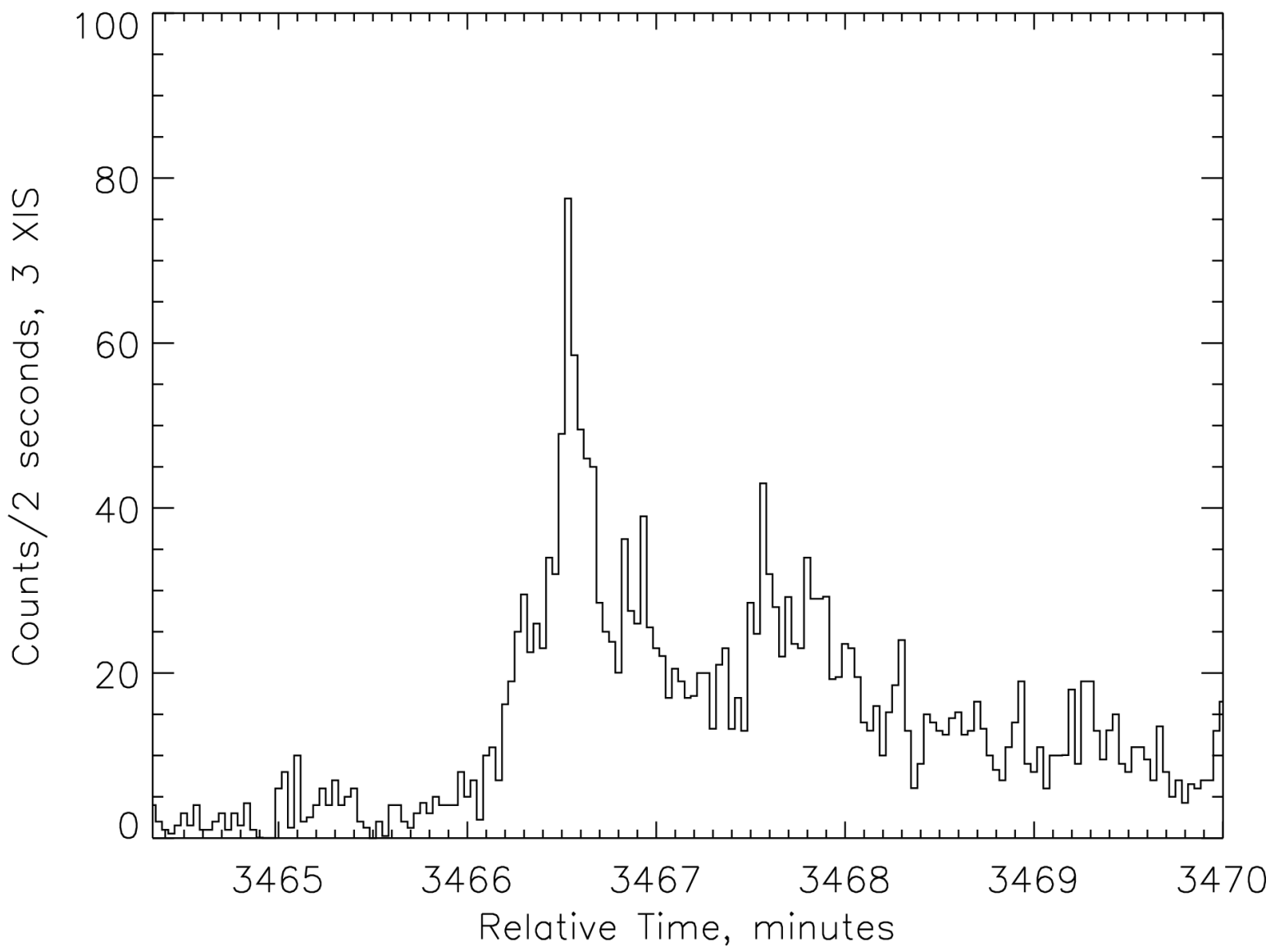

**Figure 3:** Lightcurve of the isolated outburst near 58 hours in Figure 1. The rise to the peak has a doubling time of 4 seconds or less; the 2-second binning is the native time resolution of the XIS instrument in 1/4 window mode.

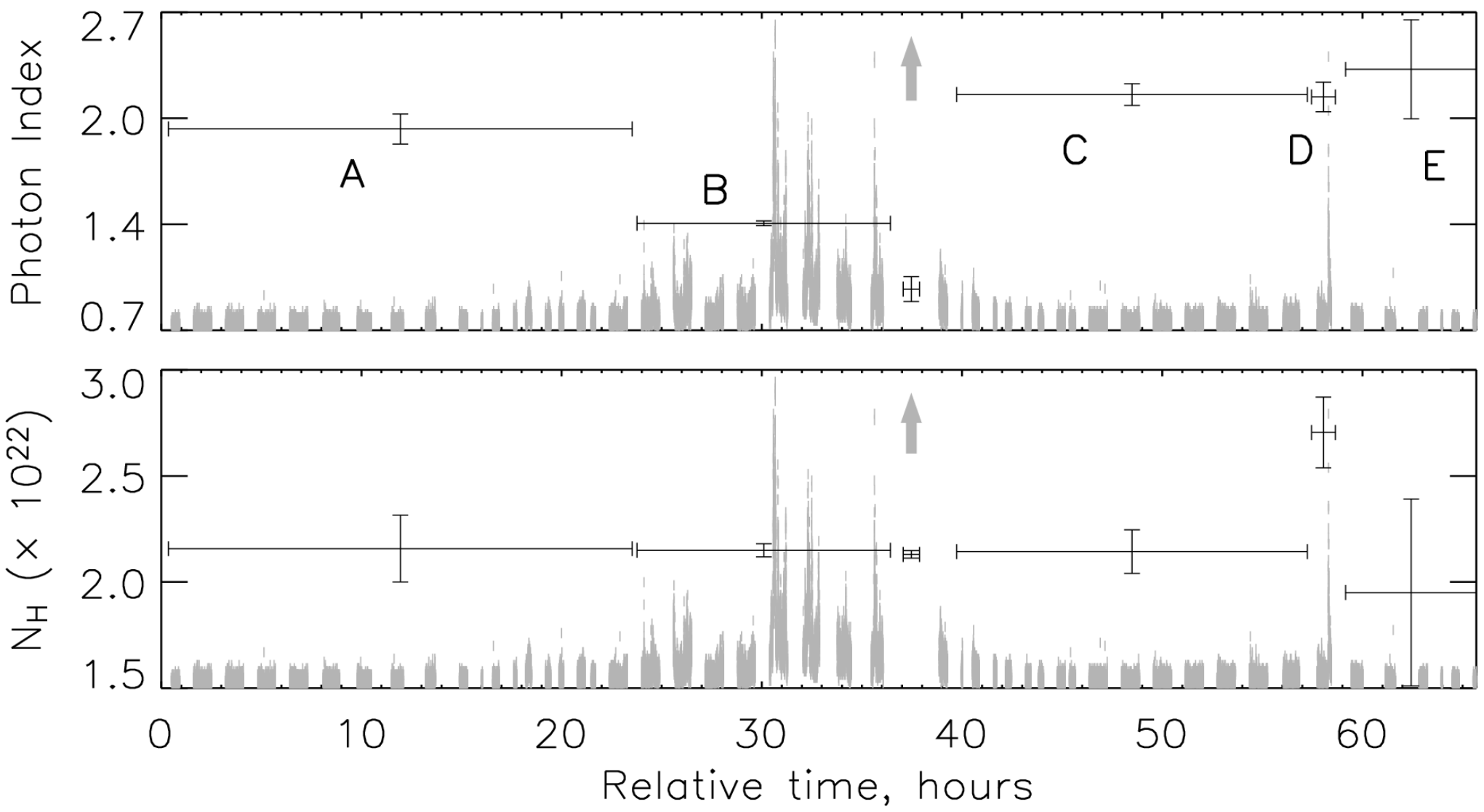

**Figure 4:** Spectral parameters $N_H$ and Γ (photon index) versus time, with the XIS lightcurve plotted in gray. The labels A through E refer to time intervals used in Table 1.

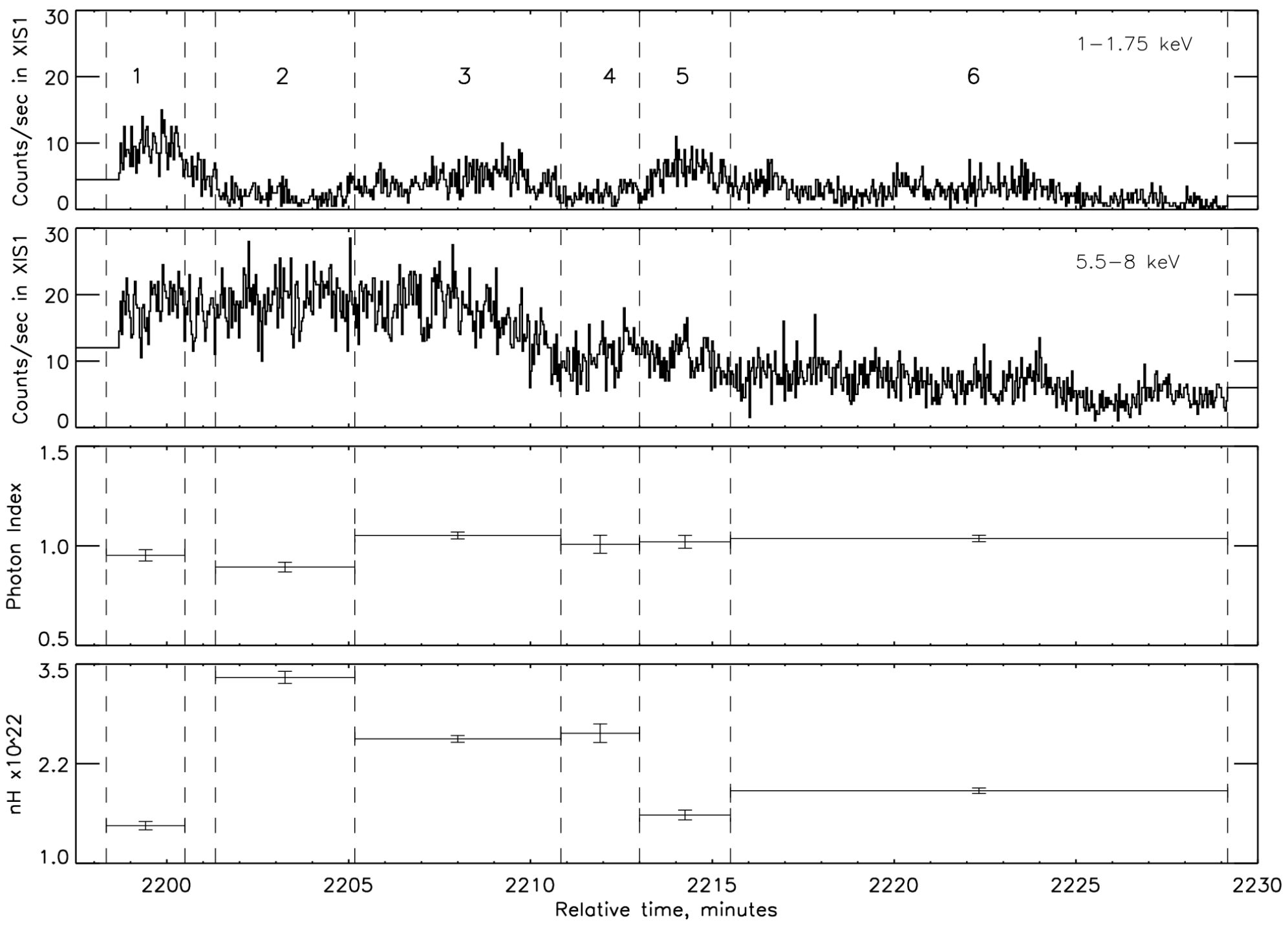

**Figure 5:** Spectral evolution of the major outburst. The top two panels are lightcurves in the energy bands 1 – 1.75 keV and 5.5 – 8 keV respectively. The bottom two panels give the spectral parameters $N_H$ and Γ.

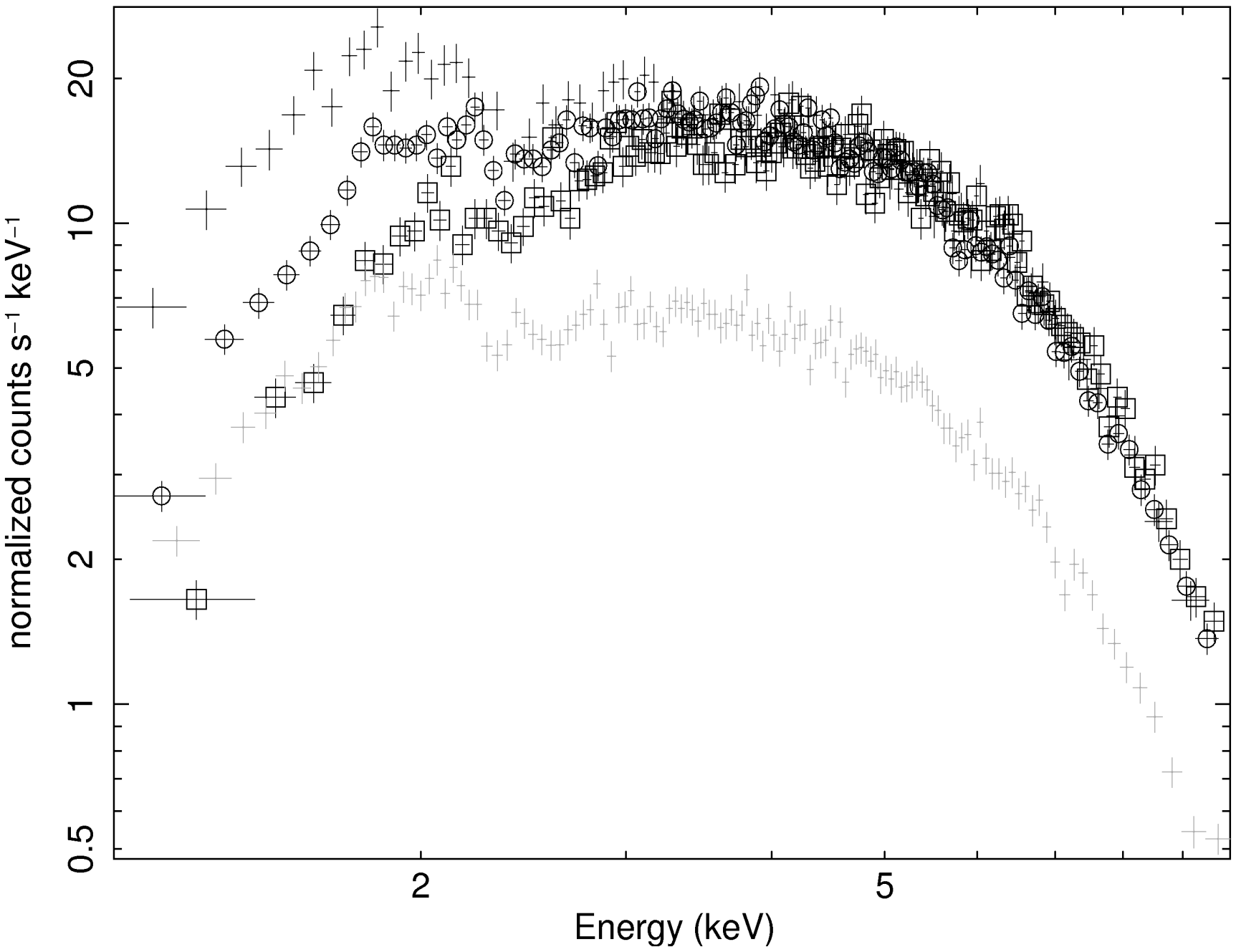

**Figure 6:** Spectra accumulated from selected time intervals during the major flare (time intervals defined in Figure 5). Intervals 1 and 6 are shown with no symbol; interval 6 is in light gray. Intervals 2 and 3 are shown with square and circular symbols, respectively.

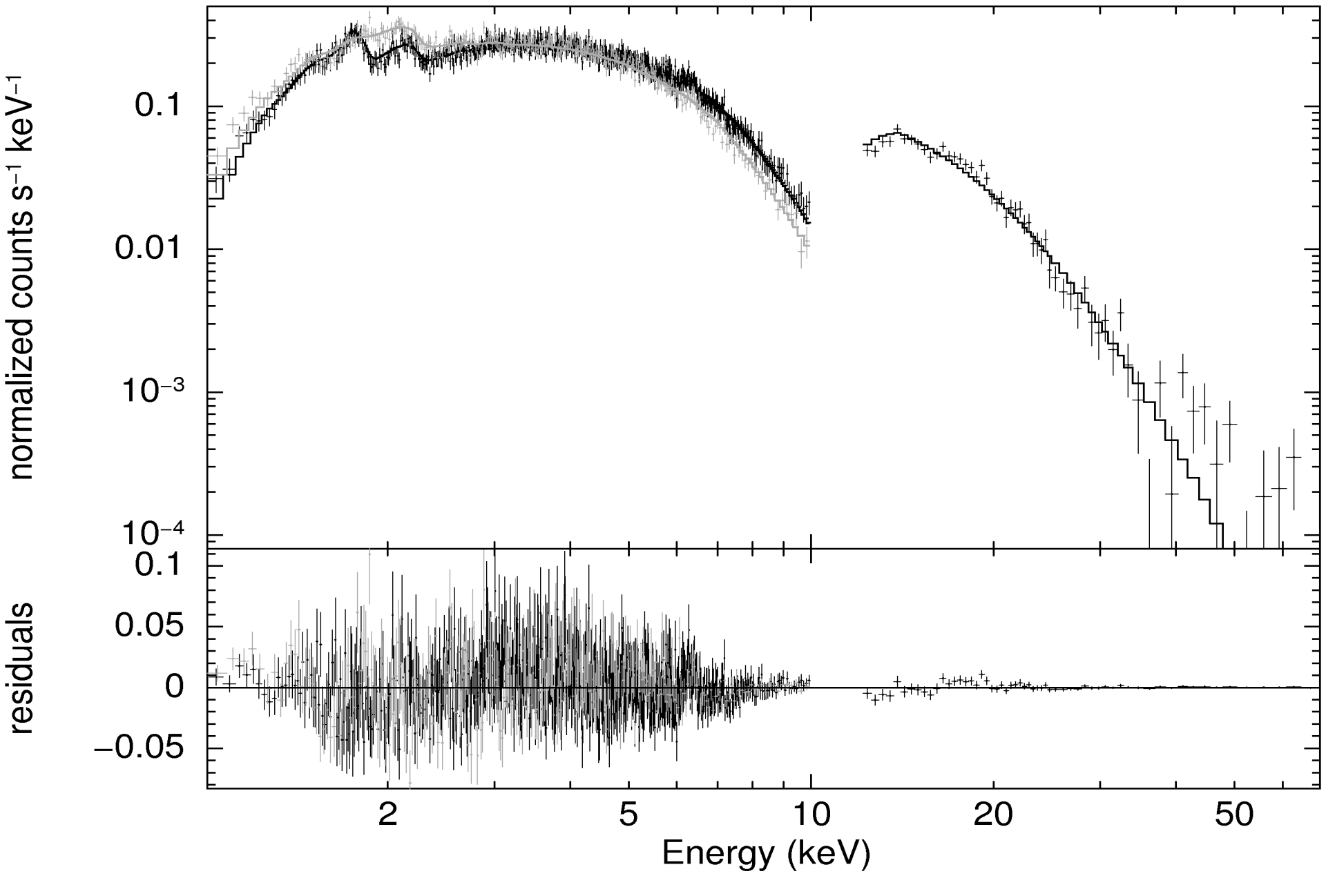

**Figure 7:** Joint fit to XIS and HXD/PIN data for period B of the observation (see Figure 4). The model is an absorbed powerlaw times an exponential cutoff (see Table 5).

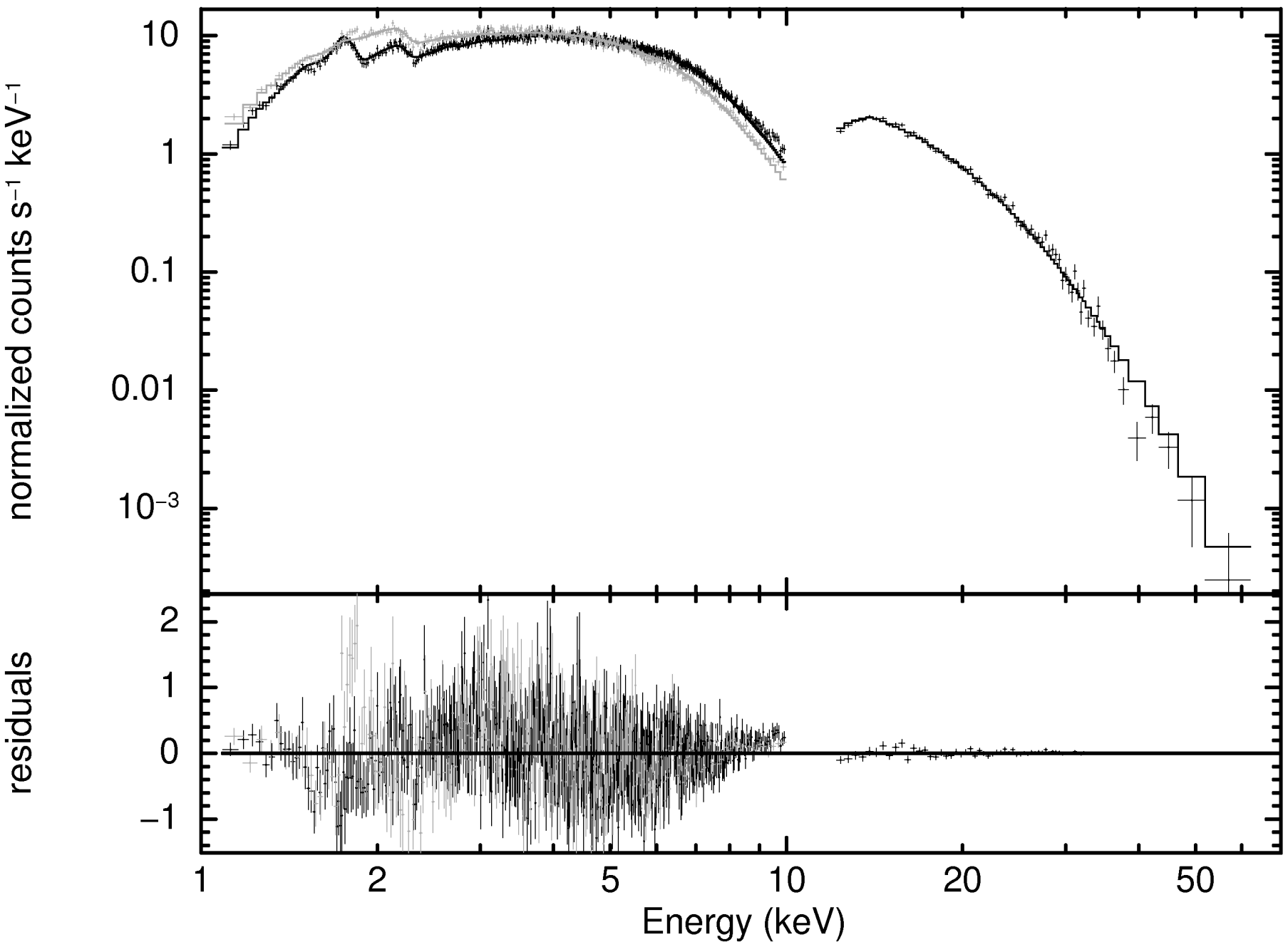

**Figure 8:** Joint fit to XIS and HXD/PIN data for the large flare (see Figures 1 and 5). The model is an absorbed powerlaw times anexponential cutoff (see text).